\documentclass[twocolumn,prb,showpacs,preprintnumbers,amsmath,amssymb,superscriptaddress]{revtex4}

\usepackage{graphicx}
\usepackage{dcolumn}
\usepackage{bm}
\usepackage{natbib}

\begin{document}

\title{Excitation spectroscopy of few-electron states in artificial diatomic molecules}

\author{T. Hatano}
\thanks{Present Address: Department of Physics, Tohoku University, Sendai 980-8578, Japan; hatano@m.tohoku.ac.jp}
\affiliation{JST, ICORP, Quantum Spin Information Project, Atsugi-shi, Kanagawa 243-0198, Japan}
\affiliation{JST, ERATO, Nuclear Spin Electronics Project, Sendai-shi, Miyagi 980-8578, Japan}

\author{Y. Tokura}
\affiliation{JST, ICORP, Quantum Spin Information Project, Atsugi-shi, Kanagawa 243-0198, Japan}
\affiliation{NTT Basic Research Laboratories, NTT Corporation, Atsugi-shi, Kanagawa 243-0198, Japan}
\affiliation{Graduate School of Pure and Applied Sciences, University of Tsukuba, Tsukuba-shi, Ibaraki, 305-8571, Japan}

\author{S. Amaha}
\affiliation{JST, ICORP, Quantum Spin Information Project, Atsugi-shi, Kanagawa 243-0198, Japan}
\affiliation{Low Temperature Physics Laboratory, RIKEN Advanced Science Institute, RIKEN, Wako-shi, Saitama 351-0198, Japan}

\author{T. Kubo}
\affiliation{JST, ICORP, Quantum Spin Information Project, Atsugi-shi, Kanagawa 243-0198, Japan}
\affiliation{NTT Basic Research Laboratories, NTT Corporation, Atsugi-shi, Kanagawa 243-0198, Japan}
\affiliation{Graduate School of Pure and Applied Sciences, University of Tsukuba, Tsukuba-shi, Ibaraki, 305-8571, Japan}

\author{S. Teraoka}
\affiliation{JST, ICORP, Quantum Spin Information Project, Atsugi-shi, Kanagawa 243-0198, Japan}
\affiliation{Department of Applied Physics, University of Tokyo, Hongo, Bunkyo-ku, Tokyo 113-8656, Japan}

\author{S. Tarucha}
\affiliation{JST, ICORP, Quantum Spin Information Project, Atsugi-shi, Kanagawa 243-0198, Japan}
\affiliation{Department of Applied Physics, University of Tokyo, Hongo, Bunkyo-ku, Tokyo 113-8656, Japan}
\affiliation{RIKEN, Center for Emergent Matter Science (CEMS), Wako-shi, Saitama 
351-0198, Japan}

\date{\today}

\begin{abstract}
We study the excitation spectroscopy of few-electron, parallel coupled double quantum dots (QDs). By applying a finite source drain voltage to a double QD (DQD), the first excited states observed in nonequilibrium charging diagrams can be classified into two kinds in terms of the total effective electron number in the DQD, assuming a core filling. When there are an odd (even) number of electrons, one (two)-electron antibonding (triplet) state is observed as the first excited state. 
On the other hand, at a larger source drain voltage we observe higher excited states, where additional single-particle excited levels are involved. Eventually, we identify the excited states with a calculation using the Hubbard model and, in particular, we elucidate the quadruplet state, which is normally forbidden by the spin blockade caused by the selection rule. 
\end{abstract}

\maketitle

Double quantum dots (QDs), which are formed by quantum mechanically coupling two QDs, are the smallest units of artificial molecules \cite{Austing,Hatano2}, and have recently been used as building blocks for spin-based quantum computing \cite{Loss}, for example, single spin qubits \cite{Koppens,Laird,Nowack,Ladriere,Obata}, singlet-triplet qubits \cite{Petta,Shulman} and two-qubit gates \cite{Brunner,Folleti}. 
These qubit operations are all performed on the two-electron states in series coupled double QDs (DQDs) by using a Pauli spin blockade \cite{Ono}.

The energy spectrum of electronic states in series DQDs have already been studied using transport measurement and charge sensing techniques. However, these techniques may not be sufficiently powerful to determine the evolution of energy levels with spin states. One reason for this is that, in a transport measurement, the elastic current only flows through the triple degenerate points, at which three different DQD charge states, namely ground states, are aligned. Accordingly, for the excited states, an inelastic cotunneling current only flows weakly near the triple degeneracy points. 
Another reason is that, as regards charge sensing, the ground and excited states cannot be distinguished. On the other hand, in parallel coupled DQDs, the current flows through all the charge states. Moreover, under a biased condition the current can also flow through the excited states as observed with single QDs \cite{Kouwenhoven}. 
In particular, the ground and excited states are well defined in vertical QDs, because a large source drain voltage can be applied thanks to the high potential barriers formed by a heterostructure \cite{Tarucha}. As a result, parallel coupled vertical DQDs may be more relevant than series coupled lateral DQDs for the excitation spectroscopy of molecular states. 

We have already reported a correlation between tunnel coupling and exchange coupling \cite{Hatanoprb} and the Aharonov-Bohm oscillations of the current flowing through one electron bonding and antibonding states \cite{Hatanoab}, by using parallel coupled vertical DQDs. 
Hitherto, two-electron states have attracted much interest in relation to exact or effective electron numbers in DQDs in relation to qubit operation, and therefore, the excited states in the other electron-number regions have not been investigated \cite{Johnson}. 
However, from such investigations, we can acquire the basic spectroscopy of DQDs (i.e. artificial molecules) and furthermore, can realize higher spin states. 
In particular, three-electron quadruplet states have attracted much attention for application to quantum computation, e.g. for DQDs, as a fast hybrid double-quantum-dot qubit \cite{Shi}, and for triple QDs, to initialize spin bits (qubits) \cite{Amaha2}. 

In this Rapid Communication we observe the ground and excited states of one- to three-electron states by using parallel coupled vertical DQDs. Assuming that the cores of the two QDs are filled, for the one-electron (two-electron) state, we observe smooth evolutions of the ground and excited states with interdot detuning, which are well explained by the anticrossing of the tunnel (exchange) coupled states. However, we observe no hybridizations of the states with different spin quantum numbers. 
For a larger bias voltage, we obtain higher excited states, and the excitation spectra observed near the triple degeneracy points are well reproduced by a numerical calculation using a Hubbard model. We identify the quadruplet state, which is normally forbidden by the spin blockade caused by the selection rule \cite{Weinmann}.

\begin{figure}
\includegraphics[width=1\columnwidth]{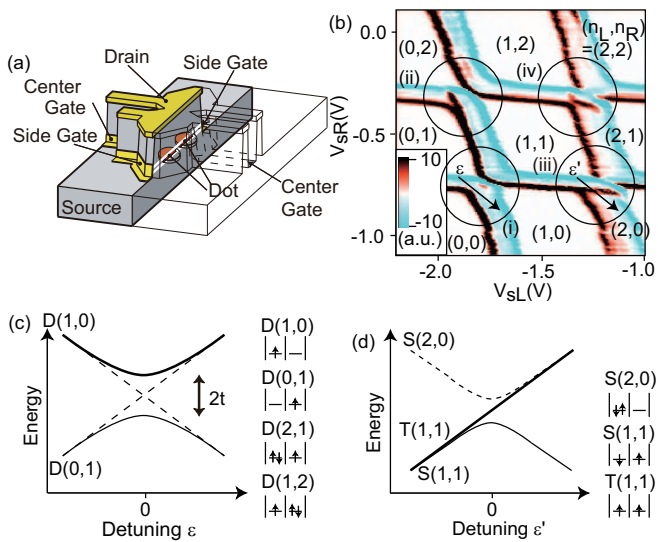}
\caption{
(a)Schematic structure of vertical double quantum dot device. (b) Average values of the transconductance $-(dI_{sd}/dV_{sL}+dI_{sd}/dV_{sR})/\sqrt{2}$ as a function of $V_{sL}$ and $V_{sR}$ at $V_{c}$=-0.55 V and $V_{sd}$=-300 $\mu$V. (c) Schematic energy diagram for one-electron doublet states in the left and right dots D(1,0) and D(0,1), corresponding to (i) in (b). The definition of detuning $\epsilon$ is shown in (b). The same schematic diagram can be shown for the doublet states D(2,1) and D(1,2), corresponding to (iv) in (b). (d) Schematic energy diagram for two singlet states S(1,1), S (2,0) and one triplet state T(1,1), corresponding to (iii) in (b). The definition of detuning $\epsilon'$ is shown in (b). Note that the dashed line is not observed because of a large excitation energy. The same schematic diagram can be shown for S(1,1), S(0,2) and T(1,1), corresponding to (ii) in (b). 
} 
\label{figure1}
\end{figure}

Figure~\ref{figure1}(a) shows a DQD device constructed in a double-barrier heterostructure, with two laterally coupled vertical QDs that have four split gates \cite{Hatano2,Hatanoprb,Hatanoab}. Two of the gates, namely the side gates, are used to vary the electron number in each QD independently, and the remaining two gates, namely the center gates, are used to tune the interdot tunnel coupling. A current, $I_{sd}$, flows in the vertical direction via the two QDs connected in parallel. The transport measurements were carried out in a dilution refrigerator at a temperature of $\sim$20 mK.

Figure~\ref{figure1}(b) shows the average value of the transconductance $-(dI_{sd}/dV_{sL} + dI_{sd}/dV_{sR})/\sqrt{2}$ as a function of the left and right side gate voltages $V_{sL}$ and $V_{sR}$ at the center gate voltage $V_c$= -0.55 V and the source drain voltage $V_{sd}$ =-300~$\mu$V. 
We observe Coulomb stripes, which indicate that the Coulomb oscillations in the linear transport regime \cite{epaps} are widened by the finite $V_{sd}$ \cite{Hatanoprb}. The black and red (blue) regions of each Coulomb stripe indicate the positive (negative) derivatives of the transconductance. The black and red regions below or to the left of each Coulomb stripe indicate the ground state, and the excited states are identified by the black and red regions inside the stripes \cite{blue}. The black circles (i) to (iv) highlight the anticrossing regions of the vertical and horizontal stripes, and we only see the first excited states in the lower left stripes. It is clear that the first excited states in (i) and (iv) repel the ground states and those in (ii) and (iii) extend straight from the vertical and horizontal ground state lines, respectively. 

Let us first consider the difference between the excited states in (i), (iv) and (ii), (iii). Here we assume that the effect of $[N_L,N_R]=[4,2]$ can be neglected as an electron-filled core \cite{Hatanoprb}, where $N_L (N_R)$ indicates the electron number in the left (right) QD. Thus, the effective electron numbers of the two dots are defined as $(n_L,n_R)=(N_L-4,N_R-2)$. We fixed $N_L$ and $N_R$ by measuring the Coulomb diamonds and charging diagrams \cite{epaps}. Note that a four-electron high-spin state obeying Hund's rule is not observed in this DQD due to the asymmetric cylindrical potential shape of the two QDs \cite{Tarucha}, and the single particle excitation energies in both QDs are larger than $|V_{sd}|$ in this region. The repulsive ground, and excited states in (i) and (iv) are assigned to the one-electron bonding and antibonding states, respectively. The ground and excited states in (i) are formed by the tunnel coupling of the $(n_L,n_R)$=(1,0), and (0,1) doublet states, which are schematically indicated by the doublet states, D(1,0) and D(0,1), respectively in Fig.~\ref{figure1}(c). The interdot energy detuning $\epsilon$ is measured from the resonance point between D(1,0) and D(0,1). As shown in Fig.~\ref{figure1}(c), the bonding and antibonding states are separated by the tunnel coupling energy $2t$ when D(1,0) and D(0,1) are aligned, namely the interdot level detuning $\epsilon$ = 0. We derive a $2t$ of $\sim$160~$\mu$eV in (i) of Fig.~\ref{figure1} (c). Note that, for (i), there is no clear bend in the Coulomb stripe in a different direction from the ground state due to the large $2t$. 

Similarly, the ground and excited states in (iv) arise from the tunnel coupling of the doublet states, D(2,1) and D(1,2), which are indicated in Fig.~\ref{figure1} (c). The orbital states involved in the interdot tunnel coupling are the same as those of D(1,0) and D(0,1), respectively, and the value of $2t$ is $\sim$~120~$\mu$eV derived from (iv) of Fig.~\ref{figure1} (b). This value is apparently smaller than that for the D(1,0) and D(0,1) states, because the electron wave function in each dot is pushed outwards to weaken the interdot tunnel coupling in (iv) with less negative $V_{sL}$ and $V_{sR}$ values than in (i). Note that we can also refer to the D(2,1) and D(1,2) states as a doublet with a hole in one of the two QDs. 

Following the above assumption, the straight excited states in the lower left stripes of (ii) and (iii) are expected to reflect the two-electron excited states in the DQD. The schematic energy diagram for (iii) is shown in Fig.~\ref{figure1}(d). S(1,1) and T(1,1) indicate the singlet and triplet states including an electron in each QD, respectively, and S(2,0) is the doubly occupied singlet state (see Fig.~\ref{figure1}(d)). The interdot energy detuning $\epsilon'$ is measured from the resonance point between S(1,1) and S(2,0). 
At a large negative value of $\epsilon'$, S(1,1) is the ground state and T(1,1) is the excited one, because of the exchange coupling energy $J$, which is given by the energy separation between singlet and triplet states. However, the two states are almost degenerate due to the small $J$ value of $\sim 4t^2/(U_{intra}-V_{inter})\sim$~30~$\mu$eV, where $2t$ is $\sim$~160~$\mu$eV, the intradot Coulomb energy $U_{intra}$ is $\sim$~1~meV and the interdot Coulomb energy $V_{inter}$ is $\sim$~0.2~meV. 
As $\epsilon'$ increases, the two singlet states are hybridized to form $\alpha{\rm S(1,1)}\pm\beta{\rm S(2,0)}$, where $\alpha$, $\beta>0$ and $\alpha^2+\beta^2=1$ and the ground state is $[{\rm S(1,1)}+{\rm S(2,0)}]/\sqrt{2}$ for the small detuning of $|\epsilon'|\sim 0$. Although S(2,0) becomes the ground state at much larger $\epsilon'$, the first excited state is T(1,1) over the entire $\epsilon'$ range. 
Therefore, comparing the ground and first excited states in (iii) of Fig.~\ref{figure1} (b) with those in Fig.~\ref{figure1} (d), the excited state can be assigned to T(1,1) \cite{neglect}. 
Note that the dashed line is not observed hereafter because of the high excitation energy. 
For (ii) in Fig.~\ref{figure1} (b), the ground and excited states can be explained by considering S(1,1), S(0,2) and T(1,1) in the same way as in Fig.~\ref{figure1} (d).

\begin{figure}
\includegraphics[width=1\columnwidth]{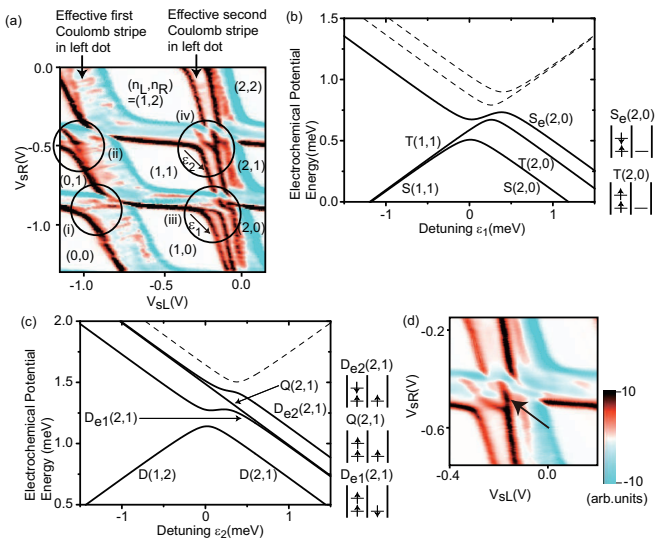}
\caption{
 (a) Average values of the transconductance $-(dI_{sd}/dV_{sL}+dI_{sd}/dV_{sR})/\sqrt{2}$ as a function of $V_{sL}$ and $V_{sR}$ at $V_c$=-0.5V, and $V_{sd}=-500\mu$V. (b) Electrochemical potential energies of two-electron states as a function of the energy detuning $\epsilon_1$ in (iii) in (a). Note that the dashed line is not observed because of a high excitation energy. (c) Electrochemical potential energies of three-electron states as a function of the energy detuning $\epsilon_2$ in (iv) in (a). Note that the dashed line is not observed because of a high excitation energy. (d) Enlargement of plot in (iv) in (a). 
}
\label{figure2}
\end{figure}

As discussed above, we can explain the excited states by assuming the [4,2] state to be the electron filled core. 
To confirm that we can generally ignore electron filled cores, we also investigate the case of a different electron filled core. 
Then, when the electron numbers in the QDs increase, the confinement energy in the QDs becomes small, and therefore, we can also observe higher excited states \cite{epaps}. 
Note that we cannot apply a $V_{sd}$ of much larger than $V_{inter}+2t$ since the nearby Coulomb stripes overlap and the structure of the excited states becomes hard to recognize.

Figure \ref{figure2}(a) shows an excitation spectrum obtained at $V_c=-0.5$~V and $V_{sd}=-500~\mu$V. More excited states can be recognized in Fig.~\ref{figure2}(a) than in Fig.~\ref{figure1}(b) \cite{epaps}. Here we also assume the [$N_L$,$N_R$]=[6,2] state to be the electron filled core and thus the effective electron number state is given by ($n_L$,$n_R$)=($N_L$-6,$N_R$-2). In (i) in Fig.~\ref{figure2}(a), the one-electron antibonding state is observed as an excited state, which is similar to (i) and (iv) in Fig.~\ref{figure1}(b), and $2t$ is estimated to be $\sim 200$~$\mu$eV. Moreover, in (ii) in Fig.~\ref{figure2}(a) the straight lines in the Coulomb stripe, i.e. T(1,1) are also obtained as in (ii) and (iii) in Fig.~\ref{figure1}(b). Note that the excited states in (i) and (ii) in Fig.~\ref{figure2}(a) are not clearer than those in Fig.~\ref{figure1} (c) due to the effect of the emitter states of the drain electrode \cite{Schmidt}. 

Here, as shown in Fig.~\ref{figure2} (a), although we can observe the excited states around the anticrossing of the two Coulomb stripes, the excited states of the single DQ are not observed in the effective first Coulomb stripe of the left QD apart from the anticrossing due to the large confinement energy. 
In contrast, we can observe two excited states in the effective second Coulomb stripe of the left QD, because the confinement energy becomes small. The first (second) excited state is the triplet (singlet) state where two electrons occupy the lowest and second lowest single-particle energy levels in the left QD with parallel (antiparallel) spins. The triplet excited state is lower than the singlet excited state due to the exchange energy. Note that the numerically calculated value of the exchange energy is approximately 15\% of $U_{intra}$ \cite{Tokura} and can be estimated to be $\sim$150~$\mu$eV, where $U_{intra}\sim$1~meV. In contrast, the confinement energy around (iv) is $\sim$300~$\mu$eV.

More intricate excited states are observed in circles (iii) and (iv) in Fig.~\ref{figure2}(a). To elucidate these complicated excited states, we calculated the electrochemical potential energies numerically using the Hubbard model, in which there are two levels in the left QD, $E_{L1}$ and $E_{L2}$, and a single level in the right QD, $E_R$. The Hamiltonian is described as follows,
\begin{eqnarray}
&\hat{H}&=\sum_{\stackrel{\scriptstyle i=L1,L2, R}{\scriptstyle \sigma=\downarrow, \uparrow}}
E_{i}c^{\dag}_{i\sigma}c_{i\sigma}
+\sum_{\stackrel{\scriptstyle i=L1,L2}{\scriptstyle  \sigma=\downarrow, \uparrow}}(tc^{\dag}_{i\sigma}c_{R\sigma}+h.c.)
\nonumber\\
&+&V_{inter}\sum_{\scriptstyle i=L1,L2}
(n_{i\downarrow}+n_{i\uparrow})(n_{R\downarrow}+n_{R\uparrow})\nonumber\\
&+& U_{intra}\sum_{\stackrel{\scriptstyle i,j=L1,L2}{\sigma, \sigma' =\downarrow, \uparrow}}n_{i\sigma}n_{i\sigma'}
+ U_{intra}\sum_{\sigma, \sigma' =\downarrow, \uparrow}n_{R\sigma}n_{R\sigma'}
\nonumber\\ 
&+&J_L\mbox{\boldmath $S$}_{L1} \cdot \mbox{\boldmath $S$}_{L2}\nonumber, 
\end{eqnarray}
where   $c^{\dag}_{i\downarrow({\uparrow})}$,  $c_{i\downarrow({\uparrow})}$ and $n_{i\downarrow({\uparrow})}$ are the electron creation, annihilation and number operators of the single-particle energy levels with a down (up) spin ($i=L1,L2$ and $R$), respectively, $J_L$ is the intradot exchange energy in the left QD (between two electrons confined at energy levels $E_{L1}$ and $E_{L2}$, respectively) , and ${\bf S}_{L1(L2)}$ is the electron spin operator of $E_{L1(L2)}$. 
The parameters are considered in relation to the experimental results: $E_{L2}-E_{L1}$=0.3 meV, $2t$=0.14 meV, $U_{intra}$=1 meV, $V_{inter}$=0.2 meV, and $J_L$=-0.15 meV. 
We construct the matrix for two- and three-electron cases, and then derive the eigenenergies by numerical exact diagonalization. The eigenenergies are used to calculate the electrochemical potentials. 

Figure \ref{figure2}(b) shows the calculated electrochemical potential for the effective two-electron region as a function of the interdot energy detuning $\epsilon_1$, corresponding to (iii) in Fig.~\ref{figure2}(a). Here $\epsilon_1$ is measured from the resonance point between S(1,1) and S(2,0). 
Adding one electron to the (1,0) ground state, we can obtain three singlet and two triplet states. 
 T(2,0)(S$_e$(2,0)) indicates a triplet (singlet) state where two electrons with parallel (antiparallel) spins occupy the lowest and second lowest single-particle energy levels in the left QD, respectively (see Fig.~\ref{figure2} (b)). 
The three singlet (two triplet) states form anticrossings as indicated by the solid lines in Fig.~\ref{figure2}(b). 
The evolution of the electrochemical potential for these states also agrees well with those in (iii) in Fig.~\ref{figure2}(a) and therefore, the excited states are now clarified. 

Similarly, to assign the states for (iv) in Fig.~\ref{figure2}(a) where three electrons contribute, the electrochemical potential energies are shown as the interdot energy detuning $\epsilon_2$ in Fig.~\ref{figure2}(c). Here $\epsilon_2$ is measured from the resonance point between D(2,1) and D(1,2). 
When one electron is added to the (1,1) ground singlet state, in addition to the ground states D(1,2) and D(2,1), two excited doublet states, i.e. D$_{e1}$(2,1) and D$_{e2}$(2,1) are realized, where the electron states in the left QD are the triplet and singlet states, respectively (see Fig.~\ref{figure2}(c)). 
In Fig~\ref{figure2}(c), the doublet states anticross, and therefore, the solid lines are identified as the ground and excited states. 

Figure~\ref{figure2}(d) shows a plot of an enlargement of the inside of (iv) in Fig.~\ref{figure2}(a). 
In addition to the excited states, D$_{e1}$(2,1) and D$_{e2}$(2,1), the straight excited state line extends vertically as indicated by the arrow in Fig.~\ref{figure2}(d), and it is impossible to explain the states using only the doublet states. However, the evolution of the electrochemical potential energy of the state is similar to the feature of the triplet states in Fig.~\ref{figure1}(d). Hence, we presume that the straight excited state is a higher spin state than D$_{e1}$(2,1), i.e. the quadruplet state, Q(2,1) for a total spin number $S=3/2$ as shown in Fig.~\ref{figure2}(c). 

According to the explanation of a conventional Coulomb blockade, the quadruplet state is forbidden by a spin blockade, which is caused by the selection rule \cite{Weinmann}.  However, T(1,1) is formed from S(1,1) via D$_{e1}$(2,1) or D$_{e2}$(2,1), and then, Q(2,1) is realized by adding an electron to T(1,1) because of the long relaxation time between the different spin states \cite{Fujisawa}. Note that it is possible to transit from T(1,1) to Q(2,1), because, in the vicinity of the center of (iv), the exchange coupling energy is estimated to be $\sim 20$~$\mu$eV and very small, where $2t$ is $\sim 130$~$\mu$eV, $U_{intra}$ is $\sim$1~meV and $V_{inter}$ is $\sim$0.2~meV. 
Consequently, Q(2,1) is indicated by a solid line in Fig.~\ref{figure2} (c). However, to clarify Q(2,1), we think that it is necessary to magnetically and electrically investigate their properties in more detail. 
Figure~2 (c), which is depicted as mentioned above, agrees well with the contents of (iv) in Fig.~\ref{figure2}(a). 


In conclusion, we have measured the excitation spectra of a few-electron, parallel coupled vertical DQD at a finite source drain voltage. 
On the assumption of core filling, the one-electron antibonding and two-electron spin triplet states were observed as the first excited states in regions where there were odd and even effective total electron numbers in the DQD, respectively. 
For a larger source drain voltage, we observed higher excited states, and elucidated their spin states by employing a numerical calculation using the Hubbard model. 
Specifically, the quadruplet state was clearly obtained. 

Finally, although we have observed the quadruplet state at a finite bias voltage, this state can easily be manipulated by pulse gate operation \cite{Fujisawa}. And we consider that measured D(2,1), D$_{e1}$(2,1) and Q(2,1) may be utilized as a quantum trit (qutrit), which is more robust than a qubit \cite{Greentree}.


We thank A. Shibatomi, S. Sasaki, and Y. Hirayama for experimental help and useful discussions. 
Part of this work is supported by a Grant-in-Aid for Young Scientists B (No. 23740248) from the Japan Society for the Promotion of Science (JSPS), a Grant-in-Aid for Scientific Research S (No. 19104007) from JSPS, a Grant-in-Aid for Scientific Research on Innovative Areas (21102003) from the Ministry of Education Culture, Sports, Science and Technology, Japan, Funding Program for World-Leading Innovative R\&D on Science and Technology (FIRST), and the IARPA project gMulti-Qubit Coherent Operationsh through Copenhagen University.



\clearpage
\widetext
\setcounter{page}{1}
\setcounter{figure}{0}
\setcounter{equation}{0}

\renewcommand{\thepage}{S\arabic{page}}
\renewcommand{\thefigure}{S\arabic{figure}}

\begin{center}

{\Large
Supplementary Material for

\vspace{0.5cm}

\baselineskip 20mm
gExcitation spectroscopy of few-electron states in artificial diatomic moleculesh
}

\vspace{0.5cm}

by

\vspace{0.3cm}

T. Hatano, Y. Tokura, S. Amaha, T. Kubo, S. Teraoka and S. Tarucha 
\end{center}


\section{Assignment of electron numbers in two dots}

\begin{figure}[hb]
\begin{center}
\includegraphics[width=100mm]{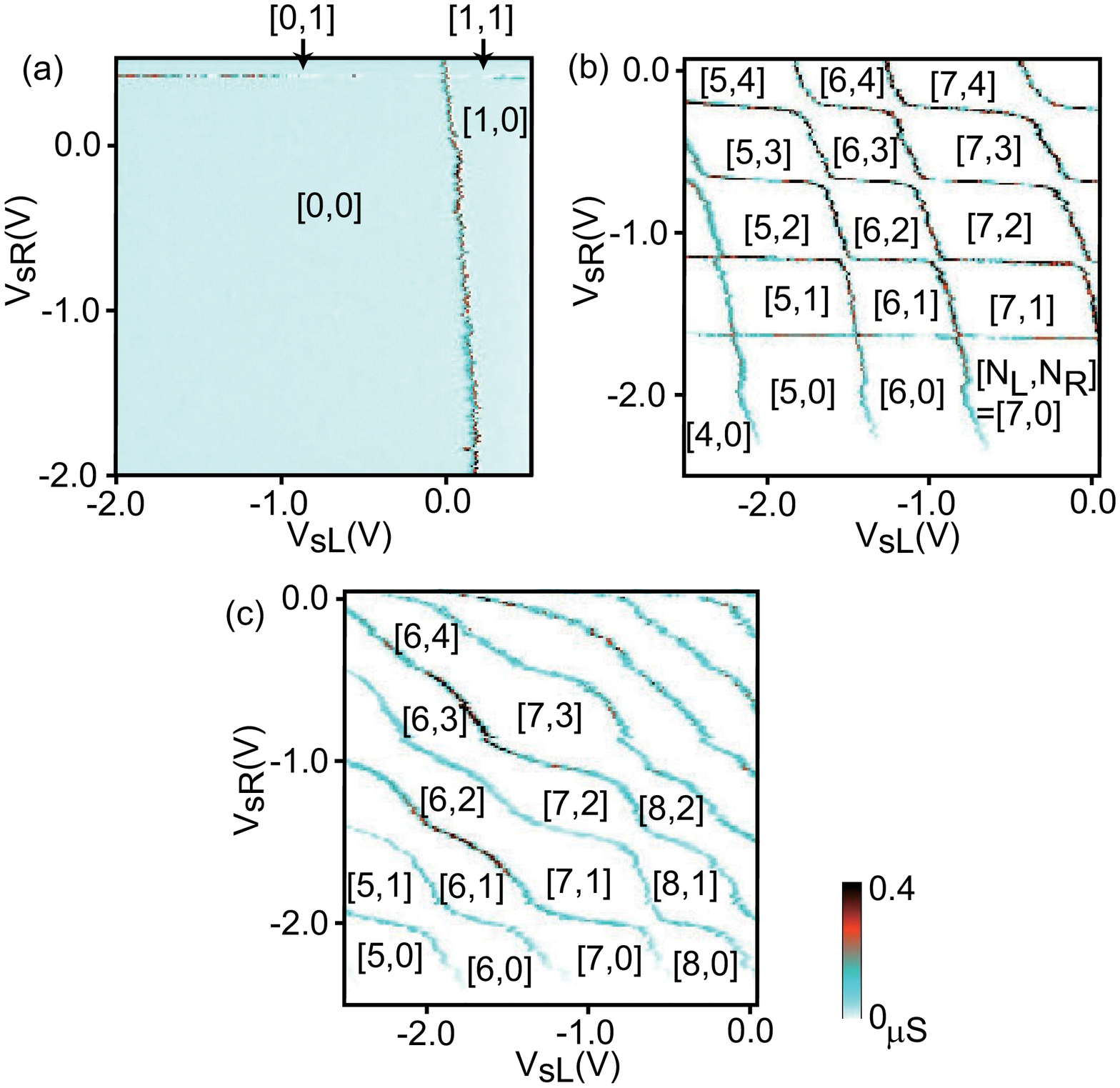}
\end{center}
\caption{
Conductance as a function of the left and right side gate voltages $V_{sL}$ and $V_{sR}$ at the source drain voltage $V_{sd}=20\mu$V for the center gate voltage (a) $V_c$=-1.8V, (b) -0.55V, (c) -0.2V. 
} 
\label{charging}
\end{figure}

In the main text, we discuss the ground and excited states for the region where the electron number in the double quantum dot (DQD) is assigned. 
Accordingly, we measured the charging diagrams as a function of the left and right side gate voltages $V_{sL}$ and $V_{sR}$ for several values of the center gate voltage $V_{c}$ to assign the electron numbers in the two QDs. The measured charging diagram at $Vc=$-1.8V is shown in Fig.~\ref{charging} (a). Here $N_L$ and $N_R$ in [ ] denote the electron numbers in the left and right QDs, respectively. At $V_{sL}\lesssim0.2$V and  $V_{sR}\lesssim0.3$V, we cannot observe Coulomb oscillations and, this indicates that there is no electron in either QD. Next, we show the charging diagram at $V_c=-0.55$V in Fig.~\ref{charging} (b). Here the position of the Coulomb oscillations shifts in the direction of the negative values of $V_{sL}$ and $V_{sR}$, when $V_c$ is made more positive. Many Coulomb oscillations can be observed in the two QDs. All the boundaries between neighboring charge states whose total electron numbers differ by one are clearly observed in the charging diagram. $N_L(N_R)$ changes by one across the vertical (horizontal) boundary lines. The vertical and horizontal lines anticross each other, indicating that the two QDs are both electrostatically and quantum mechanically coupled. 
From Figs.~\ref{charging}(a) and (b), we can assign $[N_L,N_R]$ in the closed regions. Note that the Coulomb oscillations in the two QDs almost cross in the region where there are very few electrons in Fig.~\ref{charging}(a) and (b); the tunnel coupling energy is almost zero. This is because the two QDs are separate due to the large negative $V_c$. 
 Furthermore, we show the charging diagram at $V_{c}=-0.2$V in Fig.~\ref{charging}(c). Although we can obtain more electron number regions than those in Fig.~\ref{charging}(b), the tunnel coupling is too strong to clarify the border of regions with the same total electron number; it is difficult to estimate the excited states in a region with a large $V_{c}$ value. Therefore, we measured the transport properties of the DQD in the region described in the main text to observe the excited states clearly.

\begin{figure}[hbt]
\begin{center}
\includegraphics[width=100mm]{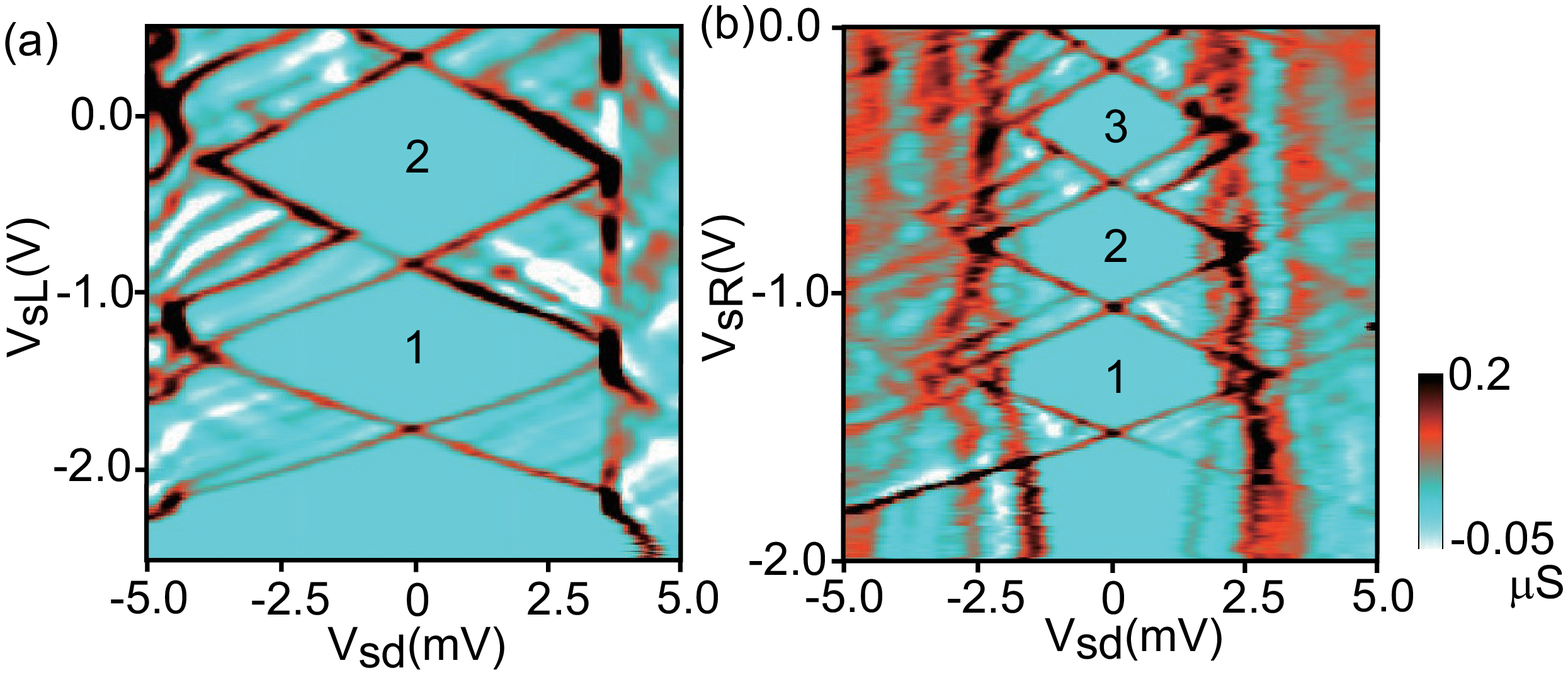}
\end{center}
\caption{
(a) Color scale plot of the numerical differential conductance $dI_{sd} /dV_{sd}$ of the left quantum dot at $V_{sR}=-1.4$V and $V_c=-1.3$V. 
 (b) Color scale plot of the numerical differential conductance $dI_{sd} /dV_{sd}$ of the right quantum dot at $V_{sL}=-2.4$V and $V_c=-0.6$V. 
The numbers in the figures indicate the numbers of electrons in the left and the right QDs. Coulomb diamonds starting from zero are observed in (a) and (b). 
} 
\label{coulomb}
\end{figure}

To confirm the zero electron regions in the two QDs more certainly, we measured the current $I_{sd}$ through only the left (right QD) as a function of $V_{sd}$ and $V_{sL(sR)}$. 
The numerical differential conductances $dI_{sd} /dV_{sd}$ of the left QD at $V_{sR}=-1.4$V and $V_c=-1.3$V are shown in Fig.~\ref{coulomb}(a) and (b), respectively. 
We cannot observe Coulomb diamonds at $V_{sL}\lesssim-1.8$V; there is no electron in the left QDs. Similarly, we also show $dI_{sd} /dV_{sd}$ for the right quantum dot at $V_{sL}=-2.4$V and $V_c=-0.6$V. 
At $V_{sR}\lesssim-1.6$V, Coulomb diamonds cannot be observed and we can confirm the zero electron region in the right QD. 
From Fig.~\ref{charging} and \ref{coulomb}, i.e. the charging diagram and Coulomb diamond measurements, we assigned the electron numbers in the two QDs.

\section{Charging diagram at small $V{sd}$}

\begin{figure}[htb]
\begin{center}
\includegraphics[width=100mm]{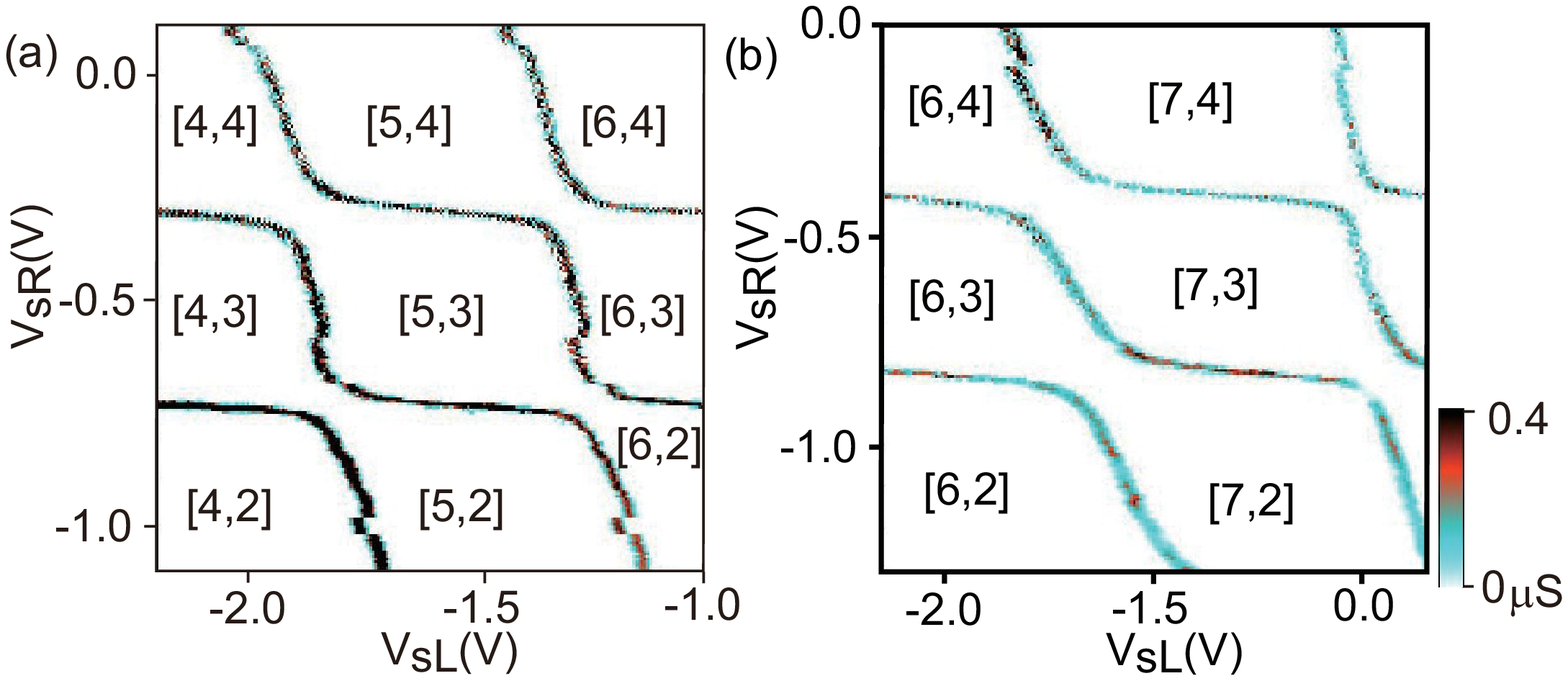}
\end{center}
\caption{
Conductance as a function of $V_{sL}$ and $V_{sR}$ at $V_{sd}=8\mu$V and (a) $V_c$=-0.55V and (b) -0.5V. The $V_{sd}$ value is 8$\mu$V. 
} 
\label{vc-055}
\end{figure}

Figures~1(b) and 2(a) in the main text were measured at finite $V_{sd}$ values. For comparison with these charging diagrams, we show the charging diagrams at $V_{sd}=8~\mu$V for (a) $V_c$=-0.55V and (b) -0.5V, which we measured for the same regions as those in Fig.~1(b) and 2(a), respectively. The electron numbers in the left and right QDs, which are indicated by the numbers in [], are assigned from Fig.\ref{charging}. As shown in Figs.~\ref{vc-055} (a) and (b), we can observe large coupling energies, which are indicated by the anticrossings of the Coulomb oscillations in the left and right QDs. Therefore, in the two regions, the ground and excited states can be clearly observed, as shown in Figs.~1(b) and 2(a) in the main text.


\begin{thebibliography}{99}
\bibitem{blue}
For the negative derivatives of the transconductance (blue regions), the ground and excited states must be observed due to electron-hole symmetry, but we cannot observe the states clearly.  This is because, when $V_{sd}$ is negative, electrons tunnel out of the DQD through the thin barrier. Therefore, the excited states are not well identified due to relaxation to the ground state.
\bibitem{neglect}
In this evaluation, since the spin-orbit interaction and hyperfine coupling are very small, the resulting singlet-triplet mixing can be neglected.
\end{thebibliography}


\end{document}